\documentclass[aps,prl,twocolumn,superscriptaddress,floatfix]{revtex4}
\usepackage{epsfig,amsmath,amssymb,color}
\begin{document}

\title{Critical properties of 1D spin-gapped fermions at the onset of magnetization}

\author{T. Vekua}
\affiliation{\mbox{Laboratoire de Physique Th\'eorique et Mod\`eles
Statistiques, CNRS, Universit\'e Paris Sud, 91405 Orsay, France}}
\author{S.I. Matveenko}
\affiliation{\mbox{Laboratoire de Physique Th\'eorique et Mod\`eles
Statistiques, CNRS, Universit\'e Paris Sud, 91405 Orsay, France}}
\affiliation{L.D. Landau Institute for Theoretical Physics, Kosygina Str. 2, 119334 Moscow, Russia} 
\author{G.V. Shlyapnikov}
\affiliation{\mbox{Laboratoire de Physique Th\'eorique et Mod\`eles
Statistiques, CNRS, Universit\'e Paris Sud, 91405 Orsay, France}}
\affiliation{\mbox {Van der Waals-Zeeman Institute,University of Amsterdam,
  Valckenierstraat 65/67, 1018 XE Amsterdam, The Netherlands}}
\date{\today}
    
\begin{abstract}
We develop an effective field theory for finding critical properties of 1D spin gapped fermions at the onset of magnetization. 
It is shown how the spin-charge interaction leads to a linear critical behavior and finite susceptibility for a wide range of models. 
We also discuss possible manifestations of spin-charge coupling in cold atomic gases.
\end{abstract}

\maketitle

Spin-charge separation is a distinctive feature of one
dimensional (1D) multicomponent fermionic/bosonic systems. In contrast to the
Fermi liquid picture, elementary excitations in 1D systems are not quasiparticles  
carrying both spin and charge, but rather spin and charge waves that propagate with
different velocities \cite{Gogolin}. This has been addressed in a number of
experimental studies and demonstrated, in particular, in experiments 
with quantum wires in semiconductors \cite{Auslaender}. 
Currently, there is a growing interest in revealing
effects of spin-charge separation in experiments with cold Fermi gases \cite{Kollath}, 
where the 1D regime has been recently achieved \cite{Esslinger,Pitaevskii}. 
     
Spin-charge separation manifests itself within effective field theory 
(bosonization) after linearizing the excitation spectrum at the Fermi points. 
However, the linearization brings in additional symmetries, for example
Lorentz invariance, and protects the system from the subtle effect of spin-charge 
interaction. The interaction between spin and charge degrees of freedom is seen in exact solutions for 
integrable systems, for instance in the Fermi Hubbard model for spin-$1/2$ fermions \cite{Frahm,Woynarovich91,Penc,brazovski}.
In this case the spin-charge coupling can also be treated by bosonization 
accounting for the curvature of the spectrum at the Fermi points \cite{brazovski}. In the presence of two gapless modes, this leads
to charge transfer by spin excitations \cite{brazovski}.

One of the achievements of bosonization is the description of the commensurate-incommensurate (C-IC) 
phase transition for spin-gapped fermions, where the gap is closed by 
a critical magnetic field and a universal square root field dependence of the
magnetization emerges at half filling\cite{Japaridze,Pokrovski}.
We show how away from half filling the spin-charge interaction leads to a linear critical behavior of magnetization 
and to a finite susceptibility at the C-IC phase transition for a fixed number of particles. The spin-charge interaction
enters the problem through the curvature of the spectrum at the Fermi points, and our effective field theory 
is applicable for a wide range of models, including continuum and extended Fermi Hubbard models with spin anisotropic 
interactions \cite{Giamarchi} and/or mass (hopping) anisotropy \cite{Cazalilla}.  In fact, for the integrable Fermi Hubbard 
model with only on-site interactions, this type of critical behavior was already obtained from the Bethe Ansatz 
\cite{Woynarovich91}. 
We give a transparent interpretation of this picture and show how the spin-charge interaction changes the behavior
of correlation functions.


We first consider a dilute strong coupling limit for spin gapped fermions and obtain the
magnetization across the C-IC transition. In this limit spin-$\uparrow$ and spin-$\downarrow$ fermions form strongly bound pairs
and the density of the thermodynamic potential is:
\begin{eqnarray}
\label{heuristic}
 {\it \Omega}=\frac{v_{\uparrow}}{2}\left[(\partial_x\phi_{\uparrow})^2 +(\partial_x\theta_{\uparrow})^2 \right] 
+\frac{v_{p}}{2}\left[(\partial_x\phi_{p})^2 +(\partial_x\theta_{p})^2 \right]\nonumber\\
-\frac{h}{2}\frac{\partial_x\phi_{\uparrow}}{\sqrt{\pi}}+W\cos{\sqrt 4\pi}\phi_{\uparrow}-\mu\frac{(\partial_x\phi_{\uparrow} +2\partial_x\phi_{p})}{\sqrt{\pi}}.
\end{eqnarray}
The fields $\partial_x\phi_p$, $\partial_x\phi_{\uparrow}$ and $\partial_x\theta_p$,  $\partial_x\theta_{\uparrow}$ represent density and current fluctuations for
the pairs and uncompensated (e.g., spin-$\uparrow$) fermions \cite{Gogolin}, $h$ is the magnetic field, the term $W$ mimics a gap for spin excitations which is closed by
a critical field
$h_{cr}$, $\mu$ is the chemical potential, and the multiple ${\cal N}=(\partial_x\phi_{\uparrow} +2\partial_x\phi_{p})/\sqrt{\pi}$ describes fluctuations of the total number of
fermions. At a fixed $\mu$ the fields $\partial_x\phi_p,\partial_x\theta_p$ and $\partial_x\phi_{\uparrow},\partial_x\theta_{\uparrow}$ are decoupled and one obtains the usual
sine-Gordon like square root dependence of magnetization on the field \cite{Japaridze,Pokrovski}: $m\sim{\sqrt{h-h_{cr}}}$, for $h\to h_{cr}+0$. 
At a constant number of particles we have a constraint $\left<{\cal N}\right>=0$, which provides coupling between the fields of spin-$\uparrow$ fermions and pairs and, as we will see,
modifies this
square root dependence to a linear one \cite{comment1}. At the critical field $h_{cr}=2\Delta$, where $2\Delta$ is equal to the binding energy of the
pairs, the low-momentum dispersion relation for spin-$\uparrow$ fermions 
is $E_{\uparrow}(k)=\sqrt{v_{\uparrow}^2k^2+\Delta^2}-\Delta \simeq v_{\uparrow}^2k^2/2\Delta$,
with $v_{\uparrow}$ being their velocity. The bound pairs disperse linearly with 
velocity $v_p\neq 0$. Minimizing ${\it \Omega}$~(\ref{heuristic}) with respect to $\partial_x\phi_p$ and $\mu$ and expressing $\mu$ through magnetization we obtain:
\begin{equation}
\label{magnetization}
2\pi v_{\uparrow}m=\sqrt{\Delta  ( h-h_{cr}-v_p \pi m)}.
\end{equation}
The appearance of the term linear in $m$ under the square root is due to resolving the constraint. Equation (\ref{magnetization}) gives $m\!=\!(h\!-\!h_{cr})/\pi v_p$ for
$h\!\rightarrow \!h_{cr}\!+\!0$, and the susceptibility is 
\begin{equation}    \label{scl}
\chi=\partial m/\partial h|_{h_{cr}}=1/\pi v_p. 
\end{equation}

We now turn to the spin and charge basis and derive from microscopic principles an asymptotically exact theory near the critical point. Taking into account the curvature of the
spectrum at the Fermi points \cite{brazovski,Haldane81}, the low-energy Hamiltonian density in the weak coupling limit can be written as:
\begin{eqnarray}
\label{nonabelian}
{\cal H}&=&\sum_{\alpha=c,s}\frac{v_{\alpha}}{2}\left[ (\partial_x\phi_{\alpha})^2/K_{\alpha}+    K_{\alpha}(\partial_x \theta_{\alpha})^2  \right ]\nonumber\\
&+&\frac{g_sv_F}{2\pi}\cos{ (\sqrt{8\pi} \phi_{s})}- h\frac{\partial_x \phi_s}{\sqrt{2\pi}}\\
&+&\frac{\sqrt{\pi}}{\sqrt{2}}\kappa
\left(\partial_x\phi_c \left[(\partial_x \phi_s)^2+(\partial_x\theta_s)^2 \right]+2\partial_x\phi_s \partial_x\theta_s\partial_x\theta_c\right)\!,\nonumber
\end{eqnarray}
where the subscripts $_c$ and $_s$ stand for the charge and spin sectors.
The fields $\partial_x \phi_c$ and $\partial_x \theta_c$ 
describe fluctuations of the charge (mass) density and
current, while $\partial_x\phi_s$ and $\partial_x\theta_s$ are fluctuations of the spin density and spin current,
with $\phi_{c,s}=( \phi_{\uparrow}\pm \phi_{\downarrow})/\sqrt{2}$ and $\theta_{c,s}=( \theta_{\uparrow}\pm \theta_{\downarrow})/\sqrt{2}$.
The Hamiltonian (\ref{nonabelian}) is applicable for a wide range of models for spin-$1/2$ fermions, including continuum and extended
Hubbard models. 
The coupling constant $g_s$, Luttinger parameters $K_{c,s}$, and deviations of the charge/spin velocities $v_{c,s}$ from the Fermi velocity $v_F$ 
depend on the Fourier transforms of the interaction potential at wavevectors $k=0$ and $k=2k_F$ \cite{Gogolin}.
For spin-gapped fermions which are SU(2) symmetric at $h<h_{cr}$, one has $g_s<0$, $K_s=1+g_s/2$, and the charge sector is gapless.

Compared to the standard bosonized Hamiltonian which is quadratic in currents and spin-charge separated, Eq.(\ref{nonabelian}) has an extra (cubic)
term \cite{Haldane81,brazovski,Teber} accounting for the curvature of the free spectrum at the Fermi points. It couples the spin and charge sectors and is 
proportional to $\kappa \equiv \partial^2E(k)/2\partial k^2|_{k_F}$.
The cubic terms within the charge sector are omitted in Eq.(\ref{nonabelian}) as they are irrelevant modifications of the linearly dispersing charge mode. Dimensionally the
curvature term is irrelevant as it has the scaling dimension equal to $3>1+1$. However, as we will argue, this term can not be dropped near the critical point
where the magnetic field closes the gap, because it couples the spin and charge sectors. \
We will show how this coupling effects the magnetic susceptibility at the critical point
when magnetization sets in.

For finding the susceptibility at a given number of particles we have to impose
a constraint: $\left< \partial_x \phi_c\right>=0$, which allows us to integrate out the charge modes.  
We calculate the contribution of the curvature term to the ground state energy at the onset of magnetization, confining ourselves 
to the terms proportional to $m^2$. For extracting these terms we write $\partial_x\phi_s=:\!\partial_x\phi_s\!:+\sqrt{2\pi }m$, with the symbol $::$ standing for the normal
ordering. Then, after integrating out charge degrees of freedom, the Euclidean action is $S_{eff}=S_s^0+S_{\kappa}$, where $S_s^0$ is the action of the sine- Gordon model at
magnetization $m\to 0$, which does not give rise to an $m^2$ contribution in the ground state energy\cite{Japaridze,Pokrovski}. 
Retaining only contributions proportional to $m^2$, the term $S_{\kappa}$ originating from the spin-charge interaction is given by:
\begin{eqnarray}
\label{effective}
&&\!\!S_{\kappa}\!\!=\!\!-\!\frac{2 m^2 \kappa^2 \pi^2 }{v^2_F}\!\!\!\int \!\!\!\!\!\sum^{i\neq j}_{i,j=0,1}\!\!\!\!\left[\partial^2_{x_iy_i} G_{c}({\bf x},{\bf y}):\
\partial_{x_i}
\phi_s({\bf x})\!::\!\partial_{y_i} \phi_s({\bf y})\!:\right.\nonumber\\
&&-\left.\partial^2_{x_iy_j} G_{c}({\bf x},{\bf y}):\!\partial_{x_i} \phi_s({\bf x})\!::\!\partial_{y_j} \phi_s({\bf y})\!:\right] \,\,d {\bf x} d{\bf y}.
\end{eqnarray}
Here $\tau=iv_Ft$ is the Euclidean time, ${\bf x}=\{x,\tau\}\equiv\{x_0,x_1\}$, and ${\bf y}=\{y,\tau'\}\equiv\{y_0,y_1\}$. For simplicity we put $K_c=1$ and $v_c=v_s=v_F$, which
does not affect our main results, and the propagator for the charge sector is 
$ G_{c}({\bf x},{\bf y})=-1/4\pi \ln((x-y)^2/a^2+(\tau-\tau')^2/a^2+1)$, where $a$ is a short distance cut-off. 

The ground state energy, from which we have to extract the $m^2$ contribution, is given by:
\begin{equation}
\label{Thevacuum}
 E_0=-\frac{1}{\int\!\!d\tau}\ln \left\{ \int D\phi_se^{-S^0_s}(1-S_{\kappa}+O(m^4)) \right\},
\end{equation}
where $\int \!\!d\tau\to \infty$  and we have written $e^{-S_{\kappa}}=1-S_{\kappa}+O(m^4)$.
As we see, Eq. (\ref{Thevacuum}) involves the calculation of the expectation value of $S_{\kappa}$ (\ref{effective}) in the vacuum of the sine- Gordon theory at $m\to 0$.

For $h=h_{cr}+0$ the vacuum of effective theory contains infinitesimally small density of solitons. 
After normal ordering with respect to the vacuum at $h=h_{cr}+0$, the $m^2$ contribution in Eq. (\ref{Thevacuum}) can be extracted using the vacuum at $h=h_{cr}-0$ due to the
relation:
\begin{equation}
\label{correl}
\left<:\!\partial_{x_i}\phi_s\!: :\!\partial_{y_j}\phi_s\!:\right>_{h_{cr}+0}\!=\!\left<\partial_{x_i}\phi_s\partial_{y_j}\phi_s\right>_{h_{cr}-0}+O(m).
\end{equation}
Eq.~(\ref{correl}) can be established from mapping the sine-Gordon model onto the massive Thirring model. Then at $K_s=1/2$, where the spin sector is equivalent
to free massive relativistic fermions, one easily gets Eq.~(\ref{correl}). For $K_s\neq 1/2$ one finds that Eq.~(\ref{correl}) holds in any order of perturbation theory in the
Thirring coupling constant.

On the other hand, for $h<h_c$ the magnetic field does not
change the states of the system and only shifts the antisoliton and soliton energies by $\sim \pm h$ so that the energy of a soliton-antisoliton pair remains the same.
Since $\partial_{x_i}\phi_s({\bf x})$  has nonzero matrix elements only between the states which can differ from each other by a certain number of 
soliton-antisoliton pairs \cite{Smirnov}, the correlation function $\left<\partial_{x_i}\phi_s\partial_{y_j}\phi_s\right>$ for $h=h_{cr}-0$ is the same as at $h=0$. At $h=0$, due
to the Euclidean invariance we have: $\left<\partial_{x_i}\phi_s\partial_{y_j}\phi_s\right>=\partial_{x_i}\partial_{y_j} G_s(r)$, where
$r=\sqrt{(x-y)^2+(\tau-\tau')^2}$ is the radial variable. 
The expectation value $\left<S_{\kappa}\right>$ is given by Eq. (\ref{effective}), with the products $\partial_{x_i}\phi_s({\bf x})\partial_{y_j}\phi_s({\bf y})$  replaced by the
corresponding correlation functions.
Then, after integrating over the angular variable,
the non-local terms of the first line of Eq.~(\ref{effective}) and those of the second line cancel each other. Accordingly, the averaged integrand in Eq.~(\ref{effective}) reduces
to: 
\!\!$$\!\lim_{a\to 0}\frac{a^2\sum_i\partial_{x_i}\partial_{y_i}G_s(r)}{2\pi(r^2+a^2)^2}\!\equiv\!\frac{\delta({\bf x}-{\bf y})}{2}\!\!\sum_i\!\!\!\!<\!\!\partial_{x_i}\phi_s({\bf
x})\partial_{y_i}\phi_s({\bf y})\!\!>\!.$$
This means that the $m^2$ contribution to $E_0$ (\ref{Thevacuum}) can be obtained by using a simplified effective action:
\begin{eqnarray}
\label{efaction}
S_{eff}\!\!\!&=&S_s^0-\frac{ m^2 \kappa^2 \pi^2   }{v^2_F}\int \!\! dx d\tau \left[(\partial_x \phi_s)^2+(\partial_{\tau}\phi_s)^2 \right]. \end{eqnarray}
One thus sees that the effect boils down  to the
renormalization of the Luttinger parameter of the spin sector (increase
of $K_s$) with $m^2$:
\begin{equation}
\label{Lutincrease}
 K_s\to K_s\left(1+{2m^2 \kappa^2 \pi^2}/{v_F^2} \right).
\end{equation}
Note that for spin-gapped fermions which are $SU(2)$ symmetric at $h<h_c$ Eq.~(\ref{Lutincrease}) encodes breaking of the $SU(2)$ symmetry.

From the rescaling of the Luttinger parameter determined by Eq.~(\ref{Lutincrease}) we obtain the following $m^2$ contribution to the ground state energy:
\begin{eqnarray}
\label{Energy}
 \Delta E_0(m^2)= \frac{ \partial{E}_0}{\partial{K_s}}\Delta K_s=\frac{2K_sm^2 \kappa^2 \pi^2 }{v_F^2} \frac{\partial{ E}_0}{\partial{K_s}}.
\end{eqnarray}
For the inverse susceptibility Eq.~(\ref{Energy}) then yields:
\begin{equation}
\label{mainformula}
v_F^2\chi^{-1}=  4K_s\kappa^2 \pi^2\partial{\cal E}_0/{\partial{K_s}},
\end{equation}
where ${\cal E}_0$ stands for the ground state energy density of the sine- Gordon model.
For $K_s\to 1$ we can follow the RG procedure\cite{Amit, Kosterlitz} in order to extract the leading universal contribution to the ground
state energy density of the sine-Gordon model. In the one-loop approach we have ${\cal E}_0=-\lambda\Delta^2/v_F$, where $\Delta$ is the soliton mass (gap), and $\lambda$ 
is a positive factor which we will fix later for the $SU(2)$ sine- Gordon case.
One- loop RG estimate of soliton mass is \cite{Amit,Kosterlitz,Dzyaloshinskii}: 
\begin{eqnarray}      \label{gap}
\!\!\Delta\!\!\simeq\!\!E_F\!\left\{\!\!\!\begin{array}{l l}
\exp\{-\frac{ \arctan \sqrt{g_s^2/(2-2K_s)^2-1} }{\sqrt{g_s^2-(2-2K_s)^2} }\}; &
  \mbox{ $ \frac{|g_s|}{(2-2K_s)} \ge 1$}\\
  \exp\{-\frac{ \mathrm {arctanh} \sqrt{1-g_s^2/(2-2K_s)^2} }{\sqrt{  (2-2K_s)^2-g_s^2} }\};& \mbox { $\frac{|g_s|}{(2-2K_s)} \le 1$}
\end{array} \right.
\end{eqnarray}
Finally, from Eq.~(\ref{mainformula})  in the vicinity of the $SU(2)$ separatrix  we obtain:
\begin{equation}
\label{susceptibility}
\chi^{-1} =\frac{4\lambda K_s\kappa^2 \pi^2 \Delta^2} {3(1-K_s)^2v_F^3}=\frac{16\lambda K_s \kappa^2 \pi^2 \Delta^2 }{3v_F^3}\ln^2{\frac{\Delta}{E_F}}
\end{equation}
up to subleading contributions. 

Equation (\ref{susceptibility}) is valid for a wide class of generic models, including those with the spin anisotropy.
Strictly speaking, the Hamiltonian (\ref{nonabelian})
requires small $g_s$ and $K_s$ close to unity. Nevertheless, one
can think of extending our results to $K_s$ away from unity, in particular to the Luther-Emery point $K_s\rightarrow 1 /2$. 
Then, it is straightforward to evaluate $\chi^{-1}$ by mapping the spin sector onto free
massive fermions, which gives:
 $\chi^{-1}\propto \kappa^2 {\partial{\cal E}_0}/{\partial{K_s}} \propto  \kappa^2  \left< (\partial_t \phi_s)^2/{v_F^2}-(\partial_x\phi_s)^2 \right> \propto (\kappa \Delta
\ln{\Delta/E_F})^{2}$. 

The most important result is that the susceptibility at the
commensurate- incommensurate phase transition stays finite if the curvature
is finite: $\kappa\neq 0$. At the onset of magnetization the susceptibility is
finite also for free fermions, where $\Delta =0$ and $\chi \sim v_F^{-1}$. However, as
we see from Eq.(\ref{susceptibility}), in
the limit of $\Delta\to 0$ the susceptibility diverges. This was previously observed for the Hubbard
model \cite{Woynarovich91} and attributed to a singular character of the zero
interaction point.

In the case of integrable Fermi Hubbard model with only on-site attractive interaction $U<0$, one has 
$1-K_s\simeq |U|/2\pi v_F$ and the result of Eq.~(\ref{susceptibility}) is similar to the Bethe Ansatz calculation in the weak coupling limit \cite{Woynarovich91}: $\chi^{-1}(|U
\to 0)=8\kappa^2\pi^3 \Delta^2/v_FU^2$. This implies that the factor $\lambda$ is equal to $3/2\pi$ on the $SU(2)$ line. 

For strong coupling the Bethe Ansatz inverse susceptibility
is given by $\chi^{-1}(|U|\to \infty)=2\pi^2\nu(1-\nu)^2/|U|$ \cite{Woynarovich91}, which at a low filling factor $\nu$
tends to our strong coupling result (\ref{scl}), with $v_p=2\pi\nu/|U|$.

We now analyze the behavior of pair and single fermion correlation functions at
the onset of magnetization ($h>h_{cr}$ and $m\to 0$) \cite{assumption}. For the Hubbard model, using explicitly the dressed charge matrix \cite{Essler}, in the presence of two
gapless modes one obtains an effective Hamiltonian density\cite{Penc,Cabra,Liu}:
\begin{equation}
\label{efftheory}
{\cal H}_{eff} = \sum_{\beta=\pm}\frac{v_{\beta}}{2}\left[  (\partial_x\phi_{\beta})^2/{K_{\beta}} +  K_{\beta}(\partial_x\theta_{\beta})^2  \right].
\end{equation}
The fields $\phi_{\pm}$ and $\theta_{\pm}$ are related to the spin and charge fields through the spin-charge mixing parameter $\xi$:
\begin{equation}
\label{transform}
\phi_+=\phi_c-\xi\phi_s,\,\theta_+=\theta_c,\,\,\,\,
\phi_-=\phi_s,\,\theta_-=\theta_s+{\xi} \theta_c,  
\end{equation}
and $v_{\pm}$, $K_{\pm}$ are the Bethe Ansatz velocities and Luttinger parameters for the $\pm$ sectors. 
For $m\rightarrow 0$ we have $v_- \propto m\to 0$, $K_-\rightarrow 1/2$ at any $U$ and $\nu$ \cite{Vekua}.
In the case of half filling ($\nu=1$) one has $K_+=1$, $\xi=0$ for all $|U|$, and there is an exact spin-charge separation so that the fields $\phi_{\pm},\theta_{\pm}$
coincide with $\phi_{c,s},\theta_{c,s}$. 
For $\nu<1$ one has
\begin{eqnarray}
K_+= 1+\frac{|U|}{2\pi v_F};\,\,\xi=\sqrt{\frac{8v_F}{|U|}}\cos{\!\left(\frac{\pi \nu}{2}\right)}\exp{\left(\!-\frac{\pi v_F}{|U|}\right)}
\end{eqnarray}
at $|U|\rightarrow 0$, with the Fermi velocity $v_F=2\sin{\pi \nu/2}$ , and $K_+=2$, $\xi=1-\nu$, $v_+=2\pi\nu/|U|$ for
$|U|\rightarrow\infty$. 
So, the parameter of spin-charge mixing, $\xi$, ranges from  $0$ to $(1-\nu)$ and monotonically increases with $|U|$.
The effective Hamiltonian (\ref{efftheory}) is obtained through the Bethe Ansatz calculation and for
the inverse susceptibility it naturally gives the exact result\cite{Woynarovich91}:
$\chi^{-1}= 2\pi v_+\xi^2/K_+$. 

Asymptotic behavior of correlation functions for the Hubbard model with a repulsive on-site interaction, in the
presence of two gapless modes (and in the presence of magnetic field) was obtained by Frahm and Korepin \cite{Frahm}. Critical exponents for the general
case have been obtained from a numerical solution of the coupled Bethe Ansatz integral equations for the dressed charge matrix. The effective Hamiltonian (\ref{efftheory}) was
constructed by Penc and S\'olyom in such a way that it reproduces the Bethe Ansatz behavior of correlation functions \cite{Penc}. This procedure of obtaining an effective
Hamiltonian was retranslated to the case
of attractive Hubbard model by using the particle hole transformation \cite{Liu}. 

The limit of $m\to 0$ allows us to derive analytical expressions for the critical exponents of the correlation functions and make a number of physical conclusions. 
For the pair correlation function from Eq.~(\ref{efftheory}) we obtain: 
\begin{equation}
\label{paircor}
\langle \psi_{\uparrow}^{\dagger}(x)\psi_{\downarrow}^{\dagger}(x) \psi_{\downarrow}(0)\psi_{\uparrow}(0) \rangle \propto
\frac{\cos{2\pi mx}}{x^{1/2+1/K_+}};\,\,\,x\rightarrow\infty,
\end{equation}
whereas for $h<h_{cr}$ it is $\sim x^{-1/K_+}$. There is a universal jump of 0.5 in the critical exponent, the result that is expected from the theory based on
spin-charge separation. However, for the single fermion Green function we find:
\begin{equation} 
\label{singlecor}
\langle \psi_{\uparrow(\downarrow)}^{\dagger}(x)\psi_{\uparrow(\downarrow)}(0) \rangle \propto
\frac{\cos{k_{F\uparrow({\downarrow})}x}}{x^{\nu_{\uparrow(\downarrow)}}};\,\,\,x\rightarrow\infty,
\end{equation}
where $k_{F\uparrow(\downarrow)}$ is the Fermi momentum of spin-up (down) fermions given by the free value. The critical exponent of the majority (spin-up) component is
$\nu_{\uparrow}=1/2+K_+/4+(1+\xi)^2/8+(1-\xi)^2/4K_+$, and for the spin-down component we obtain 
$\nu_{\downarrow}=\nu_{\uparrow}+(1/K_+-1/2)\xi>\nu_{\uparrow}$.
The presence of an additional spin-charge mixing term $\sim \xi$ in the critical
exponent of the single fermion Green function suggests that
$\nu_{\uparrow}<\nu_{\downarrow}$ even in the limit of $m\to 0$, which is a
clear signature of spin-charge coupling.
Persistence of spin- charge coupling down to $m\to 0$ limit was recently observed numerically \cite{Rizzi}.
The difference $\nu_{\downarrow}-\nu_{\uparrow}$
increases with $|U|$ for weak coupling, reaches its maximum in the regime of intermediate coupling, and then decreases with 
increasing $|U|$ in the strong coupling regime. Thus, the
effect of spin-charge mixing is the most pronounced at an intermediate coupling strength. 

In conclusion, we showed that the curvature couples
spin and charge modes for $m\to 0$ and changes critical properties of 1D spin gapped fermions at the onset of magnetization. 
Two-component Fermi gas in a 1D optical lattice is well suited for revealing spin-charge separation or
observing spin-charge coupling, especially in a box potential where $\nu$ is coordinate independent. 
Periodic modulations of the box size can only excite in-phase oscillations of the two components (charge oscillations), 
and they will not excite out-of-phase oscillations (spin mode) at half filling where exact spin-charge separation holds.
In contrast, for a significantly smaller filling factor the excitation of these modes will be provided by spin-charge coupling. 

We are grateful to A. Tsvelik, P. Wiegmann, T. Giamarchi, and G. Orso for fruitful discussions and acknowledge hospitality and support of Institut Henri Poincar\'e during the
workshop "Quantum Gases" where part of this work has been done.  
The work was also supported by the IFRAF Institute, by
ANR (grants 05-BLAN-0205 and 06-NANO-014-01), by the QUDEDIS program of ESF, and by the Dutch Foundation FOM.  
LPTMS is a mixed research unit No. 8626 of CNRS and Universit\'e Paris Sud. T.V. acknowledges GNSF grant No. N $06_-81_-4_-100$. 



\begin{thebibliography}{10}
\bibitem{comment1} In the presence of exact spin-charge separation, the chemical potential at a fixed particle number is field
independent. Thus, one can put $\mu=0$ across the transition, which gives the square root dependence $m\propto\sqrt{h-h_{cr}}$.
\bibitem{assumption} We expect that for $m\neq 0$ we have a two component Luttinger liquid, although in the limit of
$m\to 0$ the velocity of linear excitations of one of the components vanishes.
\end{thebibliography}
\end{document}